\documentstyle [12pt]{article}
\newcommand{\be}{\begin{equation}}
\newcommand{\ee}{\end{equation}}
\newcommand{\bea}{\begin{eqnarray}}
\newcommand{\ena}{\end{eqnarray}}
\newcommand{\vs}[1]{\rule[- #1 mm]{0mm}{#1 mm}}

\def\Tr   {{\rm Tr}}
\def\Pt   {{\rm P}_T}
\def\Pl   {{\rm P}_L}
\def\Re   {{\rm Re}}
\def\Im   {{\rm Im}}

\def\Ps   {\rlap/P}

\def\ie   {{\it i.e.}}
\def\(    {\left( }    \def\)   {\right) }
\def\[    {\left[}    \def\]   {\right] }
\def\txt #1 {\qquad {\rm #1} \qquad}
\def\lsim{\; \raise0.3ex\hbox{$<$\kern-0.75em\raise-1.1ex\hbox{$\sim$}}\; }
\def\gsim{\; \raise0.3ex\hbox{$>$\kern-0.75em\raise-1.1ex\hbox{$\sim$}}\; }
\def\@versim#1#2{\lower0.2ex\vbox{\baselineskip\z@skip\lineskip\z@skip
  \lineskiplimit\z@\ialign{$\m@th#1\hfil##\hfil$\crcr#2\crcr\sim\crcr}}}

\begin{document}
\renewcommand{\thefootnote}{\fnsymbol{footnote}}
\newpage
\setcounter{page}{0}

\vs{30}

\begin{center}
{\LARGE {\bf{ Photon Propagation in Dense Media}}}\\
\vspace{0.8cm}
{\large T. Altherr, E. Petitgirard, T. del R\'\i o Gaztelurrutia}\\[0.4cm]
{\em{Laboratoire de Physique Th\'eorique}}
{\small E}N{\large S}{\Large L}{\large A}P{\small P}
\footnote{URA 14-36 du CNRS, associ\'ee \`a l'E.N.S. de Lyon, et au L.A.P.P.
d'Annecy-le-Vieux.}\\
{\em B.P. 110, F-74941 Annecy-le-Vieux Cedex, France}
\\[0.8cm]
\end{center}
\vs{2}

\centerline{ \bf{Abstract}}
Using thermal field theory, we derive simple analytic expressions for the
spectral density of photons in degenerate QED plasmas, without assuming the
usual non or ultra-relativistic limit. We recover the standard results in
both cases. Although very similar in ultra-relativistic plasmas, transverse
and longitudinal excitations behave very differently as the electron
Fermi momentum decreases.
\vfill

\rightline{{\small E}N{\large S}{\Large L}{\large A}P{\small P}-A-412/92}
\rightline{December 92}

\renewcommand{\thefootnote}{\arabic{footnote}}
\setcounter{footnote}{0}
\newpage
Cores of white dwarves and red giant stars are typical examples of
degenerate QED plasmas, where the Fermi energy is much greater than the
temperature so that the Fermi distribution function can be well
approximated as a step function. The vast majority of the
literature describes the particle scatterings in these systems in the
non-relativistic limit \cite{Raf0}. But the electron Fermi
momentum, $k_F$, is typically of the same order as the electron mass.
On the other hand, it is well known that the properties of light inside
matter quite differ between the non and the ultra-relativistic limit
\cite{Raf0,Sit,AK}.
Indeed, at low densities, longitudinal time-like excitations oscillate at
fixed frequency $\omega=\omega_0$, where $\omega_0=e^2 N_e/m_e$ is the
plasmon frequency \cite{Sit}. This is not so at very high densities,
where the dispersion relations acquire a much more complicated form
\cite{AK}. Also, space-like excitations seem very different between
the two cases.

Thermal Field Theory is a unique tool for describing a system of multi
interacting particles in thermal equilibrium at a fixed temperature $T$ and
chemical potential $\mu$ \cite{Kap,LvW}. With the recent developments in
the high temperature limit \cite{BP}, this theory is now settled on a sound
basis.
Concerning degenerate plasmas, $i.e.$, $\mu\gg T$, there exists some work
related with the emission of hypothetical particles such as axions, but
dealing essentially with ultra-relativistic plasmas \cite{AK}.
We have been able to use the same methods in the general case where no
approximation is made on the electron mass. The analytic expressions turn
out to be very simple and easy to manipulate, and this is the main
motivation for presenting our results here, separately from the
applications we have in mind, like plasmon decay and axion emission
\cite{Raf0,AK,Plasmon,Bra,AKa}.

The essential mathematical object we shall manipulate is the photon
polarization tensor, together with its finite density corrections, from
which one can derive the photon spectral density.
For illustration, we use the imaginary time formalism \cite{Kap} (our
results can be derived in the real-time formalism too \cite{LvW}).
The calculations are slightly complicated by the fact that we consider a
massive fermion \cite{Toi}. However it turns out, as we show below,
that the so-called ``hard-loop'' approximation \cite{BP} can be
taken in a consistent way, leading to simple expressions. This happens when
the photon momentum is much smaller than the scales associated with the
electron, \ie, the electron chemical potential $\mu$ and the Fermi momentum
$k_F$ (here we adopt the notation of a relativistic chemical potential,
$\mu=\sqrt{k_F^2+m^2}$).

The one loop contribution to the photon polarization tensor is
\be
\Pi^{\mu\nu}(q_0,q)=- e^2 T \sum_n \int {d^3 p\over (2\pi )^3}
{ \Tr ~\gamma^{\mu}(\Ps -m)\gamma^{\nu}(\Ps'-m)\over \( P^2-m^2\)
\( P'^2-m^2\) }
,\ee
where $p_0 = i\omega_n +\mu$,  $\omega_n=\pi T(2n+1)$ and $P'=P-Q$.

The tensorial structure is usually decomposed into transverse and longitudinal
modes \cite{Wel}:
\be
\Pi^{\mu\nu} = \Pi_T \Pt^{\mu\nu} + \Pi_L \Pl^{\mu\nu}
,\ee
and, although thermal field theory is Lorentz-covariant, a preference is
given to the plasma rest frame, where $\Pt$ and $\Pl$ are given by
\be\begin{array}{ccc} \displaystyle
\Pt^{00}=0 \qquad ;&    \Pt^{0i}=0 \qquad ;&  \Pt^{ij}=
-\delta^{ij}+q^iq^j/q^2 ;\\
\Pl^{00}= -q^2/Q^2 \qquad ;&  \Pl^{0i}=-q_0q_i/Q^2 \qquad ;&
\Pl^{ij}=-(q_0^2/Q^2)(q^i q^j/q^2) .\end{array}
\ee

Let us now give some details about the calculation of the transverse
component. After performing the energy sums and taking the $T=0$ limit we get
\bea
\Pi_T(q_0,q) &=& -4e^2 \int {d^3p\over (2\pi)^3} \left\{ {1\over 2E_p}
\[ 1-\theta(-\mu-E_p) - \theta(\mu-E_p) \] \right.\nonumber\\
 & +& {p^2 \sin^2 \theta - {1\over 2}Q^2
\over 4E_p E_p'}\[ {1\over E_p'-E_p+q_0} \( \theta(\mu-E_p) -
\theta(\mu-E_p') \) \right. \nonumber\\
&   +& {1\over E_p-E_p'+q_0} \( \theta(-\mu-E_p')-
\theta(-\mu-E_p) \)  \nonumber \\
& +& {1\over E_p'+E_p-q_0} \( -1 + \theta(\mu-E_p)+
\theta(-\mu-E_p') \) \nonumber\\
& -& \left.\left.{1\over E_p'+E_p+q_0} \( 1- \theta(\mu-E_p') -
\theta(-\mu-E_p) \)  \] \right\}
.\ena
Note that in the degenerate limit $\mu\gg T$, the Fermi-Dirac distribution
functions can be approximated as step functions which greatly simplifies
the energy integrations.

\bigskip

{}From now on we will ignore the $\mu$-independent part of the polarization
tensor (absorbed in the renormalization and subleading). Also we restrict our
study to the $\mu>0$ case. We therefore obtain
\bea
\Pi_T(q_0,q) &=& 4e^2 \int {d^3p\over (2\pi)^3} \left\{ {1\over 2E_p}
\theta(\mu-E_p) - {p^2 \sin^2 \theta - {1\over 2}Q^2\over 4E_p E_p'}
\right.\nonumber\\
&&\left.\times\[ {\theta(\mu-E_p) - \theta(\mu-E_p') \over E_p'-E_p+q_0}
               + { \theta(\mu-E_p)  \over E_p'+E_p-q_0}
               + {\theta(\mu-E_p') \over E_p'+E_p+q_0} \] \right\}
.\ena
We make our approximations at this level. In an equivalent way to the hard
thermal loop approximation of Braaten and Pisarski, we study the case where
the external momenta $q$ and $q_0$ are much smaller than the Fermi
momentum. In the ultrarelativistic limit this reduces to the usual
$q,q_0 \ll \mu$. The most important contribution to $\Pi$ comes from
the region where $p$ is of the order of $k_F$. Then the following
approximation will be valid
\bea
E_p-E_p'& = &\sqrt{{\bf p}^2+m^2} - \sqrt{({\bf p-q})^2+m^2}
     \simeq  \sqrt{p^2+m^2} - \sqrt{p^2+m^2-2pq\cos \theta}\nonumber \\
&\simeq &{pq\cos \theta\over \sqrt{p^2+m^2}} =
{pq\cos\theta\over E_p} \label{cuatro}
.\ena
In the same limit, we can ignore the $Q^2$ term in eq.~(5) and the difference
between $\theta$-functions in eq.~(5) can be written as
\be
\theta(\mu-E_p) -\theta(\mu-E_p')\simeq (E_p-E_p')\delta (\mu-E_p)
.\ee
Neglecting higher orders in $q/\mu$ we obtain the final result
\be
\Pi_T (q_0,q) = {e^2\over 2\pi^2} {k_F^3\over \mu}\[ \( {\mu q_0\over k_F q}
\) ^2 + {1\over 2} {\mu q_0\over k_F q} \( 1-\( {\mu q_0\over k_F q}\)
^2\) \ln {\mu q_0 + k_F q \over  \mu q_0 - k_F q }\] .\label{cinco}
\ee

The calculation of $\Pi_L$ can be performed in an identical manner. We obtain
\be
\Pi_L (q_0,q) = {e^2\over \pi^2} { \mu k_F} \( 1-{q_0^2\over q^2}\) \[
1- {1\over 2} {\mu q_0\over k_F q}
 \ln {\mu q_0 + k_F q \over  \mu q_0 - k_F q }\]
.\label{seis} \ee
As can be seen from eqs.~(8) and (9), one-loop corrections start to be
important when $q,q_0\lsim e\sqrt{\mu k_F}$, and one should then use a
resummed photon propagator \cite{BP}.

\bigskip
The dispersion relations are simply obtained from the expressions above.
In the ultrarelativistic limit, the usual relations are recovered
\cite{AK,Toi}.
In the non-relativistic case, the dispersion relations are usually obtained
from kinetic theory \cite{Sit} and it is interesting to see how
we recover them using the methods of quantum field theory.
Throughout this paper we have considered a relativistic expression for $\mu$,
that is $\mu= m+ \mu_F$ where $\mu_F $ is the standard Fermi energy.
Therefore in the non-relativistic regime, $k_F \ll \mu$, and the mass is
the highest scale we have.
In a usual dispersion relation $q$ will always be smaller than $q_0$, and
we can thus conclude that $q k_F \ll \mu q_0$. This observation allows us to
approximate the logarithms in eqs.~(8) and (9). We find the following limiting
values for the polarization tensor
\bea
\Pi_T(q_0,q) &=& \omega_0^2\[ 1+ {1\over 5} \( {q k_F\over q_0\mu}\) ^2 \] ;\\
\Pi_L(q_0,q) &=& \omega_0^2 \( 1-{q^2\over q_0^2}\) \[
1 + {3\over 5} \( {q k_F\over q_0\mu}\) ^2\]
,\ena
where we have introduced the plasmon frequency
\be
  \omega_0^2 = {e^2\over 3\pi^2}{k_F^3\over \mu} = {e^2 N_e\over \mu}
,\ee
which is just the relativistic generalization of the usual formula.
Then, the dispersion relations are deduced by solving the pole
equation $Q^2-\Re\Pi(q_0,q)=0$ and we obtain
\bea
q_0^2& =& \omega_0^2 + \( 1+{1\over 5} v_F^2 \) q^2  \txt{when} q\ll \omega_0
\nonumber ;\\
q_0^2& =&\( 1+{1\over 5} v_F^2 \)  \omega_0^2 + q^2  \txt{when} q\gg \omega_0
,\ena
for the transverse oscillations and
\be
q_0^2 = \omega_0^2 + {3\over 5} v_F^2  q^2  \txt{when} q\ll \omega_0
,\ee
for the longitudinal.
These results agree perfectly with those obtained using kinetic theory
\cite{Sit}. We see that the transverse oscillations have particle-like
dispersion relations while the frequency of longitudinal oscillations
is almost independent of the momentum.

In Fig.~1, we have plotted the dispersion relations for a typical value of
the Fermi momentum encountered in the plasma core of red giant and white
dwarf stars. As these two systems are just between the
non and the ultra relativistic limits, the relations shown in Fig.~1 are
those to be used, for instance, in the plasmon decay process \cite{Plasmon}.

Whether it is more convenient to use thermal field theory
or kinetic theory to derive these expressions for QED is probably a matter of
taste. However, this is not the case for QCD which is
a non-abelian gauge theory and where one is forced to use our method.
As a matter of fact, the results for the gluon dispersion relations in a
degenerate quark-gluon plasma are identical to those presented here, apart
for some color factors. One should remember, however, that the quarks are
mostly ultra-relativistic as the critical density for the transition from
hadronic matter to quark-gluon plasma is much higher than the quark
masses.

\bigskip
In the calculation of emision rates of dense systems, not only the
dispersion relations, but also the imaginary part of the polarization
tensor are of interest \cite{AK}. This imaginary part gives rise to Landau
damping like contributions \cite{Pis}. From the general expressions given above
(eqs.~(8) and (9)) we see that both the longitudinal and the transverse
components develop an imaginary part whenever $q_0$ is smaller
than $(k_F/\mu)q$:
\bea
\Im \Pi_T(q_0,q) &=& -{e^2\over 4\pi} k_F^2  {q_0\over q}
                        \( 1-\( {\mu q_0\over k_F q}\) ^2\)
                           \theta(k_F q - \mu q_0) ;\\
\Im \Pi_L(q_0,q) &=& {e^2\over 2\pi} \mu^2  {q_0\over q}
                           \( 1-{q_0^2\over q^2}\)
                           \theta(k_F q - \mu q_0)
.\ena
Notice that the scales are different for the two modes.

The spectral density for space-like excitations
\be
\rho(q_0,q) = -{1\over \pi} {\Im \Pi(q_0,q) \over (Q^2-\Re\Pi(q_0,q))^2 +
              (\Im\Pi(q_0,q))^2}
,\ee
is plotted in Fig.~2 for a Fermi momentum $k_F=400$ keV. The static
limit is worth emphasizing
\bea
\Re \Pi_T(q_0=0,q\to 0) &=& 0  ;\\
\Re \Pi_L(q_0=0,q\to 0) &=& {e^2\over \pi^2} \mu k_F = k_D^2
,\ena
where $k_D$ is the inverse Debye length corresponding to the well-known
phenomenon of screening of static electric fields \cite{Sit}. Again,
the expression
given here is the relativistic generalization of the usual formula.
Transverse excitations are not screened for any range of $k_F$ and $\mu$.

Despite this screening effect, one can see in the different figures presented
here that the longitudinal spectral density has a much broader spectrum
than the transverse one, at least in the non-relativistic limit. This
originates from the different scales associated to the two modes.
As a consequence, the scattering of particles will be {\it a priori} favored
through the exchange of longitudinal photons \cite{note}. Our result
also illuminates the ``form factor'' used by several authors \cite{Raf1}.
In the language of kinetic theory, this form factor is related to the
electric permittivity $\epsilon_L$ by $F=(q^2/k_D^2)(1-1/\epsilon_L)$ while
in the language of thermal field theory the spectral density is
$\rho_L=(1/q^2)\Im(1/\epsilon_L)$ (in the static limit).
Despite the fact that the calculations are very different in the
two approaches, we can make some simple comparisons as the scales used in
the problem are the same.
In the static limit the photon propagator is just $1/(q^2+k_D^2)$ and
one would naively guess that the cross section of a particle interacting
with a charged target through the exchange of a longitudinal photon would
behave as $1/k_D^4$ \cite{Raf1}. However, as $\Im\Pi_L\sim k_D^2$ for values of
$q_0\simeq k_F q/\mu$ (which is a static point), the overall behavior is
rather like $1/k_D^2$, which is the result found in \cite{Raf1}.
We find a justification for another point: in his initial work
\cite{Raf1}, Raffelt did not take into account the damping part of the
polarization tensor into the denominator of the propagator, obtaining
a form factor $F=q^2/(q^2+k_D^2)$. However, it is clear from eq.~(17) that
if $\Im\Pi_L\sim k_D^2$ then it must be taken into account in
the denominator. This point was indeed realized by Raffelt and Seckel
\cite{RS}. The obvious conclusion is an additionnal suppression of the
scattering rates.

In sharp contrast, these considerations do not apply in the ultra-relativistic
limit. Magnetic and longitudinal scales become identical and both modes
are relevant when one considers a scattering rate. Furthermore, one
should not forget that static transverse photons are not screened \cite{AK}.

\bigskip
In conclusion, using thermal field theory, we have obtained simple analytic
expressions for the polarization tensor at a finite chemical potential $\mu$
for all range of the Fermi momentum. The relativistic generalization of the
plasmon frequency is $\omega_0^2=e^2 N_e/\mu$ and of the Debye mass
$k_D^2=(e^2/\pi^2)\mu k_F$ (these relations can also be used when the fermion
is a proton or an ion).
In agreement with the physical picture, we have found that all collective
effects (quasiparticle modes and Landau damping) occur when the photon
momentum $q,q_0$ is much smaller than the Debye mass $k_D$. As we have also
imposed the condition $q,q_0\ll k_F$ (hard loop approximation), these effects
are well described by our formulae only when $e\mu\ll k_F$, which is the
case for densities $\rho \gsim 10^5$ g/cm$^3$. In any case,
in a degenerate plasma, the photon cannot have an energy much higher than
the temperature, either because of a Boltzmann suppression factor, or
because of Pauli-blocking effects which favor the
soft photons exchange \cite{AK}. As $\mu,k_F\gg T$, one can see that the
expressions presented here (eqs.~(8) and (9)) are rather general and will
be of practical use, especially in stellar systems where neither the non
nor the ultra relativistic approximation is accurate.

\bigskip
\bigskip
{\noindent\large\bf Acknowledgements}\\
T. R. G. acknowledges the Department of Education, Universities and Research
of the Basque Goverment for financial support.

\newpage
{\large\bf Figure Caption\\[0.5cm]}
\begin{description}
\item[Fig.~1] Dispersion relations for time-like excitations for a Fermi
momentum $k_F=400$ keV. The solid line represents the transverse mode and the
dashed line the longitudinal mode.\\
\item[Fig.~2a] The spectral density for transverse space-like excitations for a
Fermi momentum $k_F=400$ keV.\\
\item[Fig.~2b] The spectral density for longitudinal space-like excitations
for a Fermi momentum $k_F=400$ keV.
\end{description}

\begin{thebibliography}{99}
\bibitem{note} At the present stage of our calculations it is however not
clear to us that transverse excitations are so much subleading compared to
longitudinal ones. After all, there is no magnetic screening in the
non-relativistic limit and this effect can give large contributions.
\end{thebibliography}
\end{document}